\documentclass[namedreferences]{SolarPhysics}
\usepackage[optionalrh,solaenum]{spr-sola-addons} 
\usepackage{graphicx}        
\usepackage{color}           
\usepackage{url}             
\usepackage[hyperindex,breaklinks]{hyperref}

\newcommand{\etal}{{\it et al.}}

\begin{document}

\begin{article}

\begin{opening}

\title{Magneto--Acoustic Energetics Study of the Seismically Active Flare of 15 February 2011}

\author{J.D.~\surname{Alvarado-G\'omez}$^{1}$\sep
	    J.C.~\surname{Buitrago-Casas}$^{1}$\sep
	    J.C.~\surname{Mart\'inez-Oliveros}$^{2}$\sep
	      C.~\surname{Lindsey}$^{4}$\sep
	      H.~\surname{Hudson}$^{2,3}$\sep
	      B.~\surname{Calvo-Mozo}$^{1}$
}

\runningauthor{Alvarado-G\'omez~\etal}
\runningtitle{Magneto--Acoustic Energetics Study}

   \institute{$^{1}$ {Observatorio Astron\'omico Nacional, Universidad Nacional de Colombia.}\\
                     email: \url{jdalvaradog@unal.edu.co} email: \url{jcbuitragoc@unal.edu.co} email: \url{bcalvom@unal.edu.co}\\ 
		$^{2}$ Space Sciences Laboratory, University of California, Berkeley, CA USA;  email: \url{oliveros@ssl.berkeley.edu},\url{hhudson@ssl.berkeley.edu}\\
		$^{3}$ Department of Physics \& Astronomy, University of Glasgow, Glasgow, Scotland, UK\\
		$^{4}$ North West Research Associates, CORA Division, Boulder, CO USA;
		email: \url{clindsey@cora.nwra.com}\\
             }

\begin{abstract}

\noindent Multi--wavelength studies of energetic solar flares with seismic emissions have revealed interesting common features between them. We studied the first GOES X--class flare of the 24th solar cycle, as detected by the Solar Dynamics Observatory (SDO). For context, seismic activity from this flare (SOL2011-02-15T01:55-X2.2, in NOAA AR 11158) has been reported in the literature \cite{2011ApJ...734L..15K,2011ApJ...741L..35Z}. Based on Dopplergram data from the Helioseismic and Magnetic Imager (HMI), we applied standard methods of local helioseismology in order to identify the seismic sources in this event.  RHESSI hard X-ray data are used to check the correlation between the location of the seismic sources and the particle precipitation sites in during the flare.  Using HMI magnetogram data, the temporal profile of fluctuations in the photospheric line-of-sight magnetic field is used to estimate the magnetic field change in the region where the seismic signal was observed.  This leads to an estimate of the work done by the Lorentz-force transient on the photosphere of the source region.  In this instance this is found to be a significant fraction of the acoustic energy in the attendant seismic emission, suggesting that Lorentz forces can contribute significantly to the generation of sunquakes.  However, there are regions in which the signature of the Lorentz-force is much stronger, but from which no significant acoustic emission emanates.

\end{abstract}


\end{opening}

\section{Introduction}
Since the discovery of a flare--induced sunquake by \inlinecite{kz1998} many other works had been written about the detection of flare--generated seismic waves in the Sun \protect\cite{zz2007,detal2006,betal2006a,metal2007,mo2007,mo2008a}. Sunquakes are defined as the observed expanding ripples on the solar photosphere induced by perturbations, driven by solar flares.  Sunquakes carry energy and momentum into the solar photosphere and interior from the photospheric footpoints of a flare.  They represent acoustic waves that travel deep into the Sun and eventually refract back to the surface after a certain time. This is observed as a close--to--circular pattern of ripples in helioseismic Dopplergrams, expanding outward from the footpoints.  The parts of the transient that travels along shallower ray paths come back to the surface nearer to the source, and sooner after the excitation, than that which penetrates deeper beneath the photosphere.  The overall picture of this process is that of an outwardly expanding patterm of surface ripples seen up to about an hour after the impulsive phase of the flare.
 
The importance of non-thermal particles (electrons and/or protons) interacting with the solar atmosphere has been already shown \protect\cite{zz2007,detal2006,metal2007,mo2007}. The study of their acceleration, propagation and precipitation have proven to be a key area in flare seismicity research. A solar model divides the flare process into several stages: acceleration, injection, propagation, trapping, precipitation and energy losses. Each of these stages has a distinctive type of electromagnetic emission and timing associated with the physical processes occurring at that stage. Therefore, studying the flare emission at different wavelengths gives important information about the physical conditions under which a sunquake can be generated. 
 
Recent studies of the acoustic emission of solar flares have opened a wide range of diagnostic possibilities and applications in helioseismology and in our understanding of these events \protect\cite{kz1998,dbl1999,dl2005}. How sunquakes are produced is still poorly understood, but, several mechanisms explaining their generation have been suggested. Some of these hypotheses rely in the energy and/or momentum that electrons may transfer to the solar interior, directly (collisions) or indirectly (back-warming radiation). These ideas are based on the close relationship evident between hard X-ray sources and the acoustic signatures \protect\cite{dl2005,2008SoPh..251..665Z}. Another hypothesis was given by \protect\inlinecite{2008ASPC..383..221H}, proposing that the magnetic field is itself the direct driver of acoustic activity during flares. In this case step--like changes of the magnetic field, as those observed by \inlinecite{2005ApJ...635..647S}, do work on the photosphere via the Lorentz force. \protect\inlinecite{2008ASPC..383..221H} estimated the magnitude of this work and found it to be comparable to the values reported in the literature.

In this work we perform a magneto--acoustic study of the first GOES X--class flare of the 24th solar cycle trying to assess if the magnetic field change during the flaring process could be an effective mechanism for the sunquake generation or if is necessary to consider more than only one mechanism acting during the flare in order to produce its sunquake phenomena.
 
\section{Observations and Analysis}
 
On 15 February 2011 the active region designated NOAA AR11158 hosted a X2.2 solar flare (SOL2011-02-15T01:55 X2.2) which began at 01:43:44 UT, reaching its maximum at 01:55:30 UT and ended at 02:27:32 UT, the first GOES X-class flare of the 24th solar cycle. To analyze this event we used Solar Dynamics Observatory/Helioseismic and Magnetic Imager (SDO/HMI) velocity and magnetic data which consist of full-disk observations in the photospheric line Fe I $6173$\AA, obtained with a cadence of 45 seconds \cite{2010SoPh..tmp..177S}. Aditionally, reconstructed images from the Reuven Ramaty High Energy Solar Spectroscopic Imager (RHESSI), observed in the $12 - 40$ and $40-100$ keV energy bands, with a spatial resolution of $\sim 1$ arcsec, were used for the hard X-ray emission of the flare (see \inlinecite{2002SoPh..210...61H}).
 
Seismic activity from this X2.2 flare was reported by \inlinecite{2011ApJ...734L..15K}. In this work, it is shown that the helioseismic waves were driven by a perturbation in the early impulsive phase, observed prior to the hard X-ray ($50-100$ keV) impulse, and probably associated with atmospheric heating by relatively low-energy electrons ($\sim 6-50$ keV) and heat flux transport. However, low-energy electrons peak at the same time as high-energy electrons (this is the ``soft-hard-soft'' phenomenon), so there is no way to distinguish them. Heat flux transport is a strange concept in this situation in general, since the impulsive phase of a flare is so nonthermal and explosive. Aditionally, the seismic source is associated with one of the perturbations located along the flare ribbons and the anisotrop of the wavefront is explained in terms of the movement of this source (see \opencite{2011ApJ...734L..15K}). Recently \inlinecite{2011ApJ...741L..35Z} analyzed the same event using acoustic holography \cite{dbl1999,dbl2000} showing the existence of a secondary weak seismic source. Also finding a spatial correlation between the seismic sources with the endpoints of a sigmoid, indicating the presence of a fluxrope in this two--ribbon flare. Based on this, \inlinecite{2011ApJ...741L..35Z} proposed a phenomenological scenario where the seismic sources are located at the footpoints of an erupting fluxrope. 

\subsection{Helioseismic analysis}

Like \inlinecite{2011ApJ...741L..35Z}, our helioseismic analysis approach for this event is based on acoustic holography.  Basically this technique corresponds to a phase-coherent reconstruction of acoustic waves observed at the solar surface into the solar interior to render ``images'' of sub-surface sources that have given rise to the surface disturbance. The main computations in holography are the ``ingression'' and ``egression'' power maps. These two quantities are estimates of the wave-field in the solar interior; the ingression is an assessment of the observed wave-field converging upon the focal point while the egression is an assessment of waves diverging from that point. 

Using this computational technique, we reconstruct egression power maps of the flaring region over different ranges of the spectrum. Figure 1 shows an spectrum of seismic emission histogram of the acoustic energy distribution derived from egression--power snapshots in the impulsive phase of the flare, because in this phase is when the sunquake mechanism is assumed to act. For the stronger source (source 1; see figure \ref{fig: 2} for source locations) most of the power is radiated in the 2.5--4.5-mHz spectrum; adding the 3-- and 4--mHz energies gives us an acoustic energy of $\sim (1.8 \pm 0.15) \times 10^{27}$ erg. The fainter source (source 2) was visible in the 6 mHz band with an estimated acoustic energy of $\sim 3\times 10^{25}$ erg.  

\begin{figure}[h!]
 \centering
 \includegraphics[bb=79 310 532 735,width=0.89\columnwidth,keepaspectratio=true]{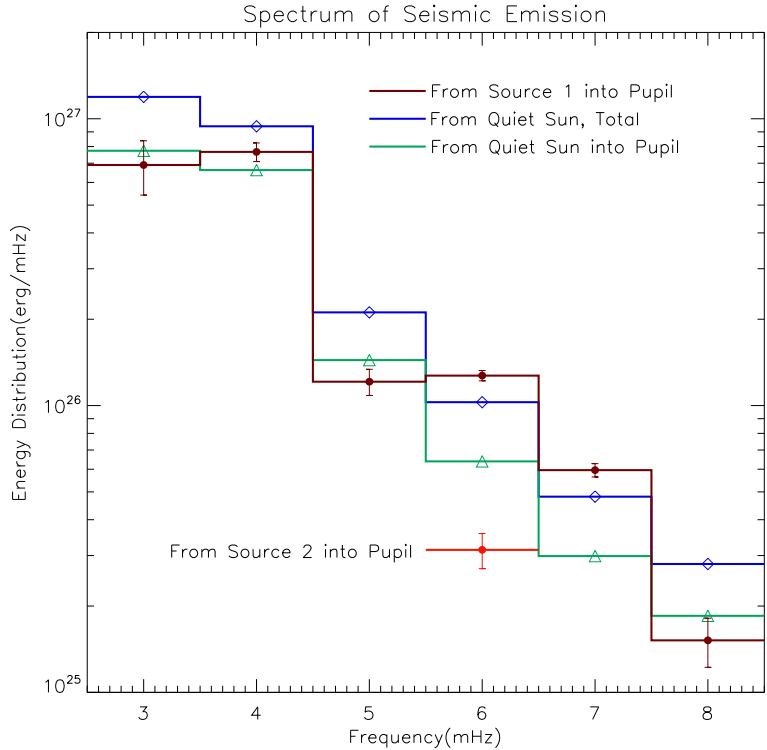}
 \caption{Energy spectrum of seismic emission from the flare of 15 February 2011 in NOAA AR11158.  The stronger source (source 1) was visible in multiple frequencies, while the fainter source (source 2) was only clearly visible in the 6 mHz band. See figure \ref{fig: 2} for source identification.}
 \label{fig: 1}
\end{figure}

\long\def\symbolfootnote[#1]#2{\begingroup\def\thefootnote{\fnsymbol{footnote}}\footnote[#1]{#2}\endgroup}

\noindent The green curve, then, represents the egression power for the limited pupil.  The blue curve is the total acoustic power.  We integrate the blue and green curves and take the ratio as an estimate of the fraction of energy emitted by acoustic sources that returns to the surface in the limited pupil of the egression computation.  These ought to be the same if all of the acoustic power emitted by the source went into the pupil and no other noise contributed to the egression-power signature. The reason the blue and green curves are as similar as can be seen in the Figure \ref{fig: 1} is that most of the power from a source of the size of that represented by the egression-power signature in Figure \ref{fig: 2} is captured by the pupil---in spite of the fact that the rest of the Sun's surface offers a much greater area\symbolfootnote[2]{This is a result of strong refraction beneath the solar surface due to a sound speed that increases rapidly with depth (see \inlinecite{lb2000}). If the sound speed were uniform in the solar interior, nearly all of an acoustic power emitted by the source would miss a small pupil such as that applied in the egression-power computations.}. 

\begin{figure}[ht!]
 \centering
 \includegraphics[bb=36 54 575 734,width=0.83\columnwidth,keepaspectratio=true]{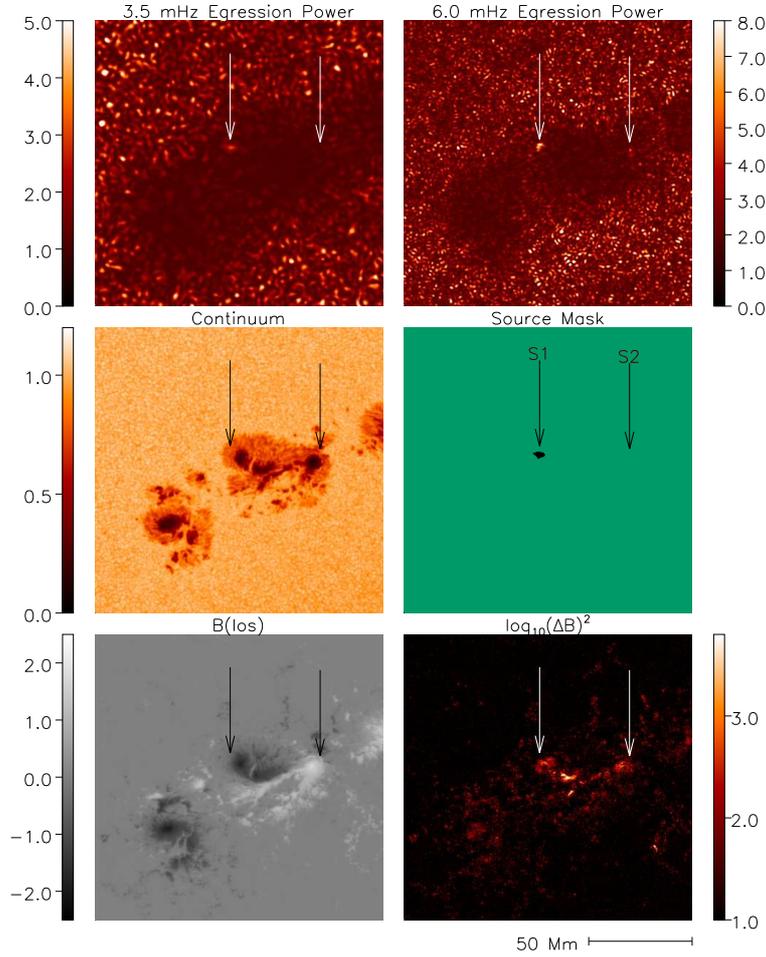}
 \caption{Acoustic source distribution for the flare of 15 February 2011 in NOAA AR11158.  Upper panels show the 2.5--4.5-mHz and 5.5--6.5-mHz egression power maps normalized to unity in the mean quiet Sun. Center left frame shows the continuum image of the flaring region while the middle right frame shows a threshold mask whose value is unity (dark) in the source region and null (green) elsewhere. Lower left shows the pre flare line-of-sight magnetic field in kG. Lower right shows the base-ten logarithm of the mean square line-of-sight magnetic variation in the 2.5--4.5~mHz spectrum during the impulsive phase of the flare.  Arrows labeled ``S1'' and ``S2'' show the locations of compact acoustic sources whose energy spectra are labeled ``Source 1'' and ``Source 2,'' respectively, in Figure \ref{fig: 1}.  The lack of a significant signature at the location labeled ``S2'' in the upper left frame signifies that excess seismic emission from Source 2, while significant in the 5.5--6.5-mHz band, is insignificant in the 2.5--4.5-mHz band.}
 \label{fig: 2}
\end{figure}

\noindent Figure 2 shows a snapshot of the 2.5--4.5 mHz egression power (upper left) during the impulsive phase of the flare cospatial with the line-of-sight magnetic field (lower left) and the base-ten logarithm of the mean square magnetic variation in the 2.5--4.5 mHz spectrum (lower right). The middle right frame presents a mask that identifies what we regard to be the main source region, obtained simply by applying a threshold to the egression-power signature of the acoustic emission evident in the upper left frame.

\subsubsection{Magnetic field variation}

\noindent According to \inlinecite{2012SoPh..277...59F}, given a change in the photospheric magnetic field, we can estimate the components of the Lorentz force resulting from magnetic restructuring. The temporal change of the force per unit area is given by 

\begin{equation}\label{eq_1}
  \frac{\partial F_z}{\partial t} = \frac{1}{8\pi}\int_{A}dA\frac{\partial}{\partial t}\left(B_{\perp}^2 - B_{z}^2\right)\mbox{ ,} 
\end{equation}

\noindent where $B_{\perp}$ and $B_z$ are the two components of the magnetic field, and the integration is made over the area $A$ where the flare-driven field changes occur, giving the total Lorentz force on the uppermost layer of the solar envelope. We can rearrange the last equation in the form

\begin{equation}\label{eq_2}
  \frac{\partial F_z}{\partial t} = \frac{1}{4\pi}\int_{A}dA\left({\bf B}_\perp\cdot\frac{\partial{\bf B}_\perp}{\partial t} - B_{z}\frac{\partial B_{z}}{\partial t}\right)\mbox{ .} 
\end{equation} 

\noindent Following the argument given by Fisher et. al (2011), we can assume that the fields are observed to change over a time interval $\Delta t$, and perform a temporal integration in the last expression in order to find the changes in the component of the Lorentz force, $\Delta F_z$:

\begin{equation}\label{eq_3}
  \Delta F_z = \frac{1}{4\pi}\int_{A}dA\left({\bf B}_\perp\cdot\Delta{\bf B}_\perp - B_{z}\Delta B_{z}\right)\mbox{ .}
\end{equation}

\noindent As we are interested in the magnetic perturbation over the photospheric surface, we consider the inward force resulting from the magnetic restructuring during the flaring process:

\begin{equation}\label{eq_4}
  \Delta F_{z,inward} = -\Delta F_z = \frac{1}{4\pi}\int_{A}dA\left(B_{z}\Delta B_{z} - {\bf B}_\perp\cdot\Delta{\bf B}_\perp\right)\mbox{ .}
\end{equation}

\noindent In order to quantify the Lorentz--force contribution to the energetics of the sunquake, some theoretical asumptions must be made taking into account the avaliable data for the estimation. First of all we will assume that $\Delta B_z$ is null, which it would have to be in the case of magnetic flux frozen into a photospheric medium  that is rigid to the extent that we cannot see a change in the geometry  of the magnetic region over the acoustic relaxation time, $\tau_{ac} \sim c/2H \sim 40$ s. Then $B_z\Delta B_z ~-~ {\bf B}_\perp\cdot\Delta{\bf B}_\perp$ is simply  $-{\bf B}_\perp\cdot\Delta{\bf B}_\perp$, which we will suppose is of the same order of magnitude as $B_{\rm los}\Delta B_{\rm los}$. This can possibly be mistaken.  For example, $\Delta{\bf B}$ could have a magnitude of five times that of $\Delta B_{\rm los}$ if $\Delta{\bf B}_\perp$ were within about a fifth of a radian of being perpendicular to the line of sight.   The probability of this would appear to be inversely proportional to the angle to within which $\Delta{\bf B}$ would have to be perpendicular to the line of sight.

What follows, then, is an estimate of $B_{\rm los}\Delta B_{\rm los}$, but the designation ``(los)'' is left off from this point on. We compute the mean value of the line-of-sight magnetic field over the mask shown in the Figure \ref{fig: 2}, to find that 

\begin{equation}\label{eq_5}
 \langle B\rangle ~=~ 500 {\rm ~G},
\end{equation}

\noindent where the angular brackets indicate the spatial average of their contents over the region identified by the source mask (see Figure \ref{fig: 2}, top right panel) . 

\begin{figure}[ht!]
 \centering
 \includegraphics[bb=79 225 532 735,width=0.83\columnwidth,keepaspectratio=true]{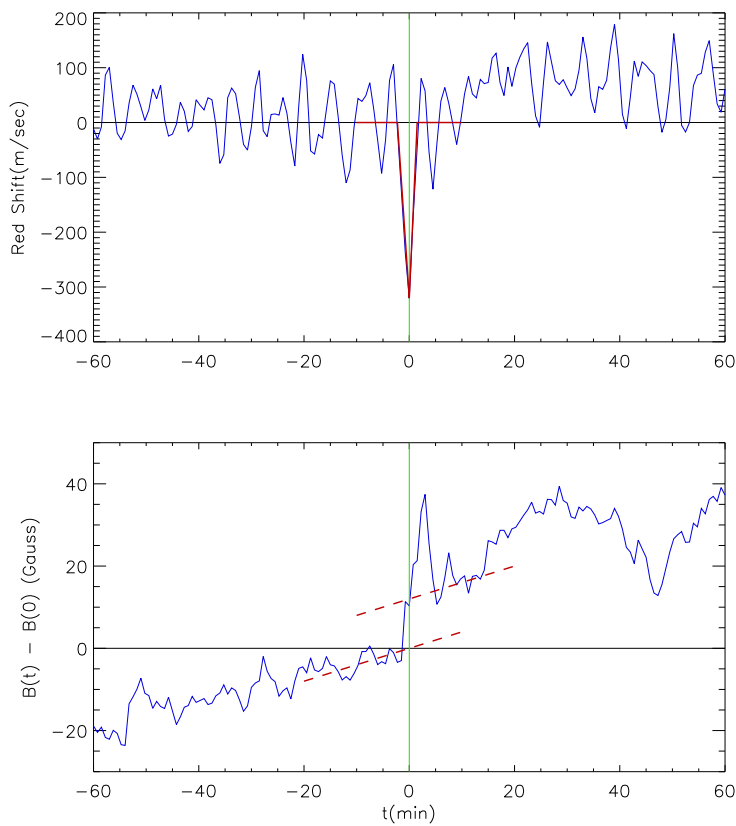}
 \caption{Doppler (upper panel) and line-of-sight magnetic field (lower) integrated over the source region defined by the 
threshold mask represented in the upper left panel of Figure \ref{fig: 2} are plotted as a function of time, $t$. Vertical green lines at time zero identify the impulsive phase of the flare. Red fang-like profile in the upper frame represents the Doppler profile of the flare uncontaminated by the 5-minute oscillations. Dashed red lines in the lower frame identify what we propose to recognize as a step-function transient component of the line-of-sight magnetic profile.}
 \label{fig: 3}
\end{figure}

\noindent Variations in the Doppler and line-of-sight-magnetic signatures spatially integrated over the source region are plotted in Figure \ref{fig: 3}. The top frame shows the Doppler signature (red shift) integrated over the source region. This shows the regular 5--minute oscillations and a significant transient behavior during the impulsive phase of the flare, which occurs at time zero. The bottom frame shows the line-of-sight magnetic field strength, $B$, integrated over the same region minus the $500$ G mean. It also shows the transient behavior of the field. We suggest that this profile, in the time frame of the impulsive phase, has an initial excursion of some 37 G that is mostly ephemeral; the excursion returns to a value that is $\sim$12 G above from the pre-flare trend, which was increasing at a rate of $0.4$ G/min, indicated by the slope of the dashed red lines in the bottom frame of Figure \ref{fig: 3}.

Now, we take the significant magnetic transient to be the step-function-like component shown by the displacement between the dashed red lines in the lower plot, the displacement between these being 12 G:

\begin{equation}\label{eq_6}
 \Delta\langle B\rangle ~=~ 12 {\rm ~G},
\end{equation}

\long\def\symbolfootnote[#1]#2{\begingroup\def\thefootnote{\fnsymbol{footnote}}\footnote[#1]{#2}\endgroup}

\noindent which is a small change compared to previous magnetic field restructuring studies during energetic flares; for instance, \inlinecite{sh2005} report, for a sample of 15 X-class solar flares, a mean value of 90 G for the magnetic field change, using Global Oscillation Network Group (GONG) magnetograms, while \inlinecite{2005JGRA..11008104Z}, using Michelson Doppler Imager (MDI) magnetograms, report an absolute change of 120 G for a GOES X4.8 class flare.     

In principle, these changes are in line with a model of magnetic-energy release that requires a ``permanent change'' in the magnetic configuration to drive a sunquake. Physically speaking, the energy released by a coronal magnetic field leaves the coronal field permanently changed in some respect, so that $B^2/8\pi$ integrated over the corona reflects the reduced magnetic energy following the flare \cite{h2000}. The magnetic virial theorem requires a significant manifestation of this changed configuration in any surface, $\cal S$, that effectively ``surrounds'' the region of reduced magnetic energy, so long as the field is force free on $\cal S$ and throughout its interior. This strongly suggests that such a change will have a manifestation in the photospheric magnetic field, through which the magnetic flux passes. Because of this, a step-function component in the photospheric magnetic signature marking the impulsive phase of the flare is of particular interest as a prospective source of work done on the solar interior acoustic field. An ephemeral change that is reversed once the impulsive phase of the flare has run its course is suspect in that it might be no more than the signature of a perturbed radiative environment during the impulsive phase. Of course, it is possible that the radiative environment itself is permanently changed after the impulsive phase of the flare. Moreover, the required change need not occur in the neighborhood of the acoustic source to satisfy the magnetic virial theorem.

In accordance with the foregoing considerations, we take the transient magnetic force density (per unit area, hence a pressure) to be\symbolfootnote[2]{Because $B_{\rm los}$ varies considerably across the source region, the mean of $B\Delta B$ can be considerably different from the product of the means used as a proxy in equation (\ref{eq_7}). The error could be of order 100\%. However, we think it is unlikely to be anything like an order of magnitude.}

\begin{equation}\label{eq_7}
 \Delta p_m ~=~ {1 \over 4\pi}\langle B\Delta B\rangle ~\sim~ \langle B\rangle \Delta\langle B\rangle ~=~ 480 ~{{\rm dyne} \over {\rm cm}^2}.
\end{equation}

\noindent Counting the number of unity-value pixels in the mask, we find an area of

\begin{equation}\label{eq_8}
 A ~=~ (105~{\rm pix}) ~dx ~dy,
\end{equation}

\noindent with pixel dimensions

\begin{equation}\label{eq_9}
 dx ~=~ dy ~=~ 5\times 10^{-4}R_\odot ~=~ 0.35~{\rm Mm},
\end{equation}

\noindent hence,

\begin{equation}\label{eq_10}
 A ~=~ 12.8~{\rm Mm}^2 ~=~ 1.3\times 10^{17}~{\rm cm}^2.
\end{equation}

\noindent This leads to a total force transient integrated over the source region of

\begin{equation}\label{eq_11}
 \Delta F_z ~=~ \Delta p_m A ~=~ 3.12\times 10^{19}~{\rm dyne}.
\end{equation}

\noindent Now, to estimate the work done on the acoustic field by this force, we need to determine how far, $\Delta z$, it depresses the photosphere in the time frame of the acoustic relaxation time, roughly 40 s in the quiet photosphere. One way to do it is to integrate over the fang-like profile plotted in red in the top panel of Figure \ref{fig: 3}, supposing that, and admitting significant errors, the displacement uncontaminated by 5--minute oscillations would look something like this. We take the area thus subtended to be

\begin{equation}\label{eq_12}
 \Delta z ~=~ {1 \over 2} ~{\rm base} \times {\rm height} ~=~ 0.5 \times 225 {\rm ~s} \times 320 {\rm ~m/s} ~=~ 36 {\rm ~km} ~=~ 3.6 \times 10^6 {\rm ~cm},
\end{equation}

\noindent where the base and height of the profile represent the duration (225 s) and velocity (320 m/s), respectively, of the photospheric depression. We now take the work done by the force over the displacement supposing that the mean imbalance of the force as the displacement progresses is half the initial transient, $\Delta F$, above; then,

\begin{equation}\label{eq_13}
 \Delta W ~=~ {1 \over 2} ~\Delta F_z ~\Delta z ~=~ 1.12\times 10^{26}\hspace{2pt}{\rm erg}.
\end{equation}

\noindent The energy estimate expressed in equation (\ref{eq_13}) is less than the $\sim (1.8 \pm 0.15) \times 10^{27}$ erg based on the helioseismic signature of the flare (source 1). Given the other relatively rapid changes in the line-of-sight magnetic signature evident in the Figure \ref{fig: 3}, the possibility ought to be considered that the force transient is greater than the step-function component. If we suppose that it is something like the full 37 G range of the excursion seen between in the time interval $-$1.0 to $+$2.5 min, for example, this would increase the estimate of $\Delta W$ proportionately, to $3.36 \times 10^{26}$ erg. This value reaches $\sim$ 18\% of the egression-power estimate for the analyzed source, representing too  considerable a fraction of it to be regarded as negligible; it must be recalled at this point that this estimation is considering only the line--of--sight component of the magnetic field, and the total energetic contribution from the magnetic field restructuring can be even higher than the value presented here.  

\subsubsection{Hard X-ray emission}

\begin{figure}[h!]
 \centering
 \includegraphics[bb=0 0 708 368,width=0.9\columnwidth,keepaspectratio=true]{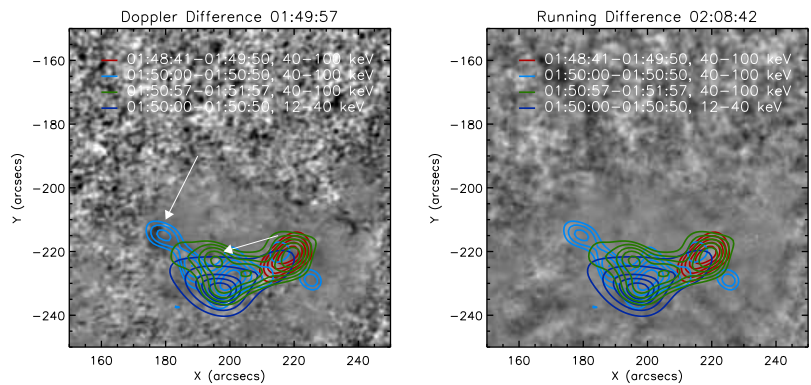}
 \caption{HMI dopplergram differences at the beginning of the impulsive phase (left) and at the sunquake wavefront detection time (right), combined with RHESSI CLEAN contour plots, with levels 50\%, 60\%, 70\%, 80\% and 90\% of the maximum detected intensity, in the $40 - 100$ and $12 - 40$ keV bands, integrated during the intervals shown. The white arrows indicate the location of the endpoints of one of the flare ribbons. The left arrow points to the center of the analyzed region.}
 \label{fig: 4}	
\end{figure}

\noindent Using RHESSI hard X-ray data we determined the spatial location of the hard X-ray footpoints in four different time intervals, as is shown in the Figure \ref{fig: 4}, during the impulsive phase where the transients in both, the magnetic and Doppler, were observed. These observations revealed a complex structure in the distribution of footpoints during this phase, suggesting multiple magnetic loop formation and movement along the flare ribbons. This process was likely to end with the relaxation of the magnetic field lines and the precipitation of particles in the endpoints of one the flare ribbons. The left white arrow indicates the center of the analyzed region (mask in the Figure \ref{fig: 2}). 

\section{Discussion}

\noindent We have studied the first GOES X--class flare of the 24th solar cycle detected by HMI/SDO, the 15 February 2011 X2.2 flare \cite{2011ApJ...734L..15K,2011ApJ...741L..35Z}, confirming the two induced sunquakes by this event.  \inlinecite{2008ASPC..383..221H} and \inlinecite{2012SoPh..277...59F} prescribe a means to estimate the contribution of the Lorentz-force transient to such a sunquake based upon transient vector magnetic signatures across the impulsive phase of the flare.  What we have undertaken here is a preliminary, order-of-magnitude estimate of the Lorentz-force contribution based on just the line-of-sight magnetic signature, pending the availability of reliable vector magnetograms from HMI.  The full contribution based on the vector-magnetic measurements could be greater or less than this estimate, but basic statistics favor an expectation that is at least somewhat greater.  Hence, if $\bf B$ and $\Delta{\bf B}$ could be regarded as isotropically distributed, then the rough estimate we derive here is most likely somewhat of an underestimate.

Our estimate of the work done by the Lorentz force contribution based on just the line-of-sight transient magnetic signature is $(1.12 \pm 0.1)\times 10^{26}$ erg, approximately 6\% of the $1.8\times 10^{27}$ erg we estimate for the total energy in the acoustic transient.  This alone is a significant fraction, and if we knew it to be no more than this, it would be dangerous to regard it as negligible for general purposes.  In fact, if we revive the proposition that the vertical component of $\Delta{\bf B}_\perp$, is null, then it should follow that

\begin{equation}\label{eq_14}
\Delta B_{\rm los} = \rho \Delta{\bf B}\cdot{\bf\hat o},
\end{equation}

\noindent where $\Delta{\bf B}$ is the componet of ${\bf B}_\perp = {\bf B}$ that lies in the local vertical plane containing the line-of-sight direction, $\bf\hat o$, and $\rho$ is the sine of the angle between $\bf\hat o$ and the vertical direction, $\bf\hat z$.  For AR11158 at the time of the flare, $\rho$ was 0.267.  Hence, under the assumption that $\Delta B$ is horizontal, it follows that the real value of the component of $\Delta B$ to which the line of sight is sensitive is $\Delta B/\rho = (12/0.267)$ G $= 45$ G. Replacing the 12 gauss estimate of equation (\ref{eq_13}) by 45 gauss increases the $10^{26}$ erg estimate to $3.75\times 10^{26}$ erg.  This is about 23\% of the $1.8\times 10^{27}$ erg acoustic energy we estimated. This remains more likely an underestimate than an overestimate, and exceeds the acoustic energy in the (5--7 mHz) where the mapping of the egression-power signature is most conspicuous (as it has been in previously detected sunquakes, see \opencite{dl2005}).  

At the same time, maps of the mean square line-of-sight magnetic transient (see lower right panel in Figure \ref{fig: 2} for this in the 2.5--4.5 mHz spectrum) show rms variation in $\Delta B_{\rm los}$ up to nearly 100 G with no correction for $\rho$ in regions far from Sources 1 and 2 and from which {\it no} significant acoustic emission emanates.  Moreover, RHESSI, the hard X-ray emission of the event show a complex structure of footpoints with multiple locations of particle precipitation during the impulsive phase, without a clear spatial correlation between these precipitation sites and the location of the acoustic sources.  Whether or not Lorentz-force transients contribute significantly to flare acoustic emission, then, appears to remain an open question, and this seems to involve some very interesting physics that has yet to be identified. 

These results also show the need for a more detailed study that utilizes high-resolution vector magnetic observations and includes the acoustics of heating of the outer atmosphere by accelerated particles. Crucial to such a study are realistic numerical simulations of the physics of prospective contributors to flare acoustic emission.

\begin{acks}
The Berkeley group was supported by NASA under contract NAS 5-98033 for RHESSI.
\end{acks}

\bibliographystyle{spr-mp-sola}
\bibliography{Biblio}

\begin{thebibliography}{22}
\ifx \bisbn   \undefined \def \bisbn  #1{ISBN #1}\fi
\ifx \binits  \undefined \def \binits#1{#1}\fi
\ifx \bauthor  \undefined \def \bauthor#1{#1}\fi
\ifx \batitle  \undefined \def \batitle#1{#1}\fi
\ifx \bjtitle  \undefined \def \bjtitle#1{\textit{#1}}\fi
\ifx \bvolume  \undefined \def \bvolume#1{\textbf{#1}}\fi
\ifx \byear  \undefined \def \byear#1{#1}\fi
\ifx \bissue  \undefined \def \bissue#1{#1}\fi
\ifx \bfpage  \undefined \def \bfpage#1{#1}\fi
\ifx \blpage  \undefined \def \blpage #1{#1}\fi
\ifx \burl  \undefined \def \burl#1{\textsf{#1}}\fi
\ifx \href  \undefined \def \href#1#2{\textsf{#2}}\fi
\ifx \doiurl  \undefined \def
  \doiurl#1{\href{http://dx.doi.org/#1}{\textsf{#1}}}\fi
\ifx \betal  \undefined \def \betal{\textit{et al.}}\fi
\ifx \binstitute  \undefined \def \binstitute#1{#1}\fi
\ifx \bctitle  \undefined \def \bctitle#1{#1}\fi
\ifx \beditor  \undefined \def \beditor#1{#1}\fi
\ifx \bpublisher  \undefined \def \bpublisher#1{#1}\fi
\ifx \bbtitle  \undefined \def \bbtitle#1{\textit{#1}}\fi
\ifx \bedition  \undefined \def \bedition#1{#1}\fi
\ifx \bseriesno  \undefined \def \bseriesno#1{\textbf{#1}}\fi
\ifx \blocation  \undefined \def \blocation#1{#1}\fi
\ifx \bsertitle  \undefined \def \bsertitle#1{\textit{#1}}\fi
\ifx \bsnm \undefined \def \bsnm#1{#1}\fi
\ifx \bsuffix \undefined \def \bsuffix#1{#1}\fi
\ifx \bparticle \undefined \def \bparticle#1{#1}\fi
\ifx \barticle \undefined \def \barticle#1{}\fi
\ifx \botherref \undefined \def \botherref#1{}\fi
\ifx \url \undefined \def \url#1{\textsf{#1}}\fi
\ifx \bchapter \undefined \def \bchapter#1{}\fi
\ifx \bbook \undefined \def \bbook#1{}\fi
\ifx \bcomment \undefined \def \bcomment#1{#1}\fi
\ifx \oauthor \undefined \def \oauthor#1{#1}\fi
\ifx \citeauthoryear \undefined \def \citeauthoryear#1{#1}\fi
\def \endbibitem {}
\ifx \bconflocation  \undefined \def \bconflocation#1{#1} \fi

\bibitem[\protect\citeauthoryear{Besliu-Ionescu
  \textit{et~al.}}{2006}]{betal2006a}
\begin{barticle}
\bauthor{\bsnm{Besliu-Ionescu}, \binits{D.}},
\bauthor{\bsnm{Donea}, \binits{A.-C.}},
\bauthor{\bsnm{Cally}, \binits{P.}},
\bauthor{\bsnm{Lindsey}, \binits{C.}}:
\byear{2006},
\batitle{Significant acoustic activity in ar10720 on january 15, 2005}.
\bjtitle{Romanian Astronomical Journal}
\bvolume{16},
\bfpage{203}.
\end{barticle}
\endbibitem

\bibitem[\protect\citeauthoryear{Donea and Lindsey}{2005}]{dl2005}
\begin{barticle}
\bauthor{\bsnm{Donea}, \binits{A.-C.}},
\bauthor{\bsnm{Lindsey}, \binits{C.}}:
\byear{2005},
\batitle{Seismic emission from the solar flares of 2003 october 28 and 29}.
\bjtitle{Astrophysical Journal}
\bvolume{630},
\bfpage{1168}\,--\,\blpage{1183}.
\end{barticle}
\endbibitem

\bibitem[\protect\citeauthoryear{Donea, Braun, and Lindsey}{1999}]{dbl1999}
\begin{barticle}
\bauthor{\bsnm{Donea}, \binits{A.-C.}},
\bauthor{\bsnm{Braun}, \binits{D.C.}},
\bauthor{\bsnm{Lindsey}, \binits{C.}}:
\byear{1999},
\batitle{Seismic images of a solar flare}.
\bjtitle{Astrophysical Journal}
\bvolume{513},
\bfpage{143}\,--\,\blpage{146}.
\end{barticle}
\endbibitem

\bibitem[\protect\citeauthoryear{Donea, Lindsey, and Braun}{2000}]{dbl2000}
\begin{barticle}
\bauthor{\bsnm{Donea}, \binits{A.}},
\bauthor{\bsnm{Lindsey}, \binits{C.}},
\bauthor{\bsnm{Braun}, \binits{D.C.}}:
\byear{2000},
\batitle{Stochastic seismic emission from acoustic glories and the quiet sun}.
\bjtitle{Solar Physics}
\bvolume{192},
\bfpage{321}\,--\,\blpage{333}.
\end{barticle}
\endbibitem

\bibitem[\protect\citeauthoryear{Donea \textit{et~al.}}{2006}]{detal2006}
\begin{barticle}
\bauthor{\bsnm{Donea}, \binits{A.-C.}},
\bauthor{\bsnm{Besliu-Ionescu}, \binits{D.}},
\bauthor{\bsnm{Cally}, \binits{P.S.}},
\bauthor{\bsnm{Lindsey}, \binits{C.}},
\bauthor{\bsnm{Zharkova}, \binits{V.V.}}:
\byear{2006},
\batitle{Seismic emission from a m9.5-class solar flare}.
\bjtitle{Solar Physics}
\bvolume{239},
\bfpage{113}\,--\,\blpage{135}.
\end{barticle}
\endbibitem

\bibitem[\protect\citeauthoryear{{Fisher}
  \textit{et~al.}}{2012}]{2012SoPh..277...59F}
\begin{barticle}
\bauthor{\bsnm{{Fisher}}, \binits{G.H.}},
\bauthor{\bsnm{{Bercik}}, \binits{D.J.}},
\bauthor{\bsnm{{Welsch}}, \binits{B.T.}},
\bauthor{\bsnm{{Hudson}}, \binits{H.S.}}:
\byear{2012},
\batitle{{Global Forces in Eruptive Solar Flares: The Lorentz Force Acting on
  the Solar Atmosphere and the Solar Interior}}.
\bjtitle{Solar Physics}
\bvolume{277},
\bfpage{59}\,--\,\blpage{76}.
\end{barticle}
\endbibitem

\bibitem[\protect\citeauthoryear{Hudson}{2000}]{h2000}
\begin{barticle}
\bauthor{\bsnm{Hudson}, \binits{H.S.}}:
\byear{2000},
\batitle{Implosions in coronal transients}.
\bjtitle{Astrophysical Journal}
\bvolume{531},
\bfpage{75}\,--\,\blpage{77}.
\end{barticle}
\endbibitem

\bibitem[\protect\citeauthoryear{{Hudson}, {Fisher}, and
  {Welsch}}{2008}]{2008ASPC..383..221H}
\begin{bchapter}
\bauthor{\bsnm{{Hudson}}, \binits{H.S.}},
\bauthor{\bsnm{{Fisher}}, \binits{G.H.}},
\bauthor{\bsnm{{Welsch}}, \binits{B.T.}}:
\byear{2008},
\bctitle{{Flare Energy and Magnetic Field Variations}}.
In: \beditor{\bsnm{{R.~Howe, R.~W.~Komm, K.~S.~Balasubramaniam, \&
  G.~J.~D.~Petrie }}} (ed.)
\bbtitle{Subsurface and Atmospheric Influences on Solar Activity},
\bsertitle{Astronomical Society of the Pacific Conference Series}
\bseriesno{383},
\bfpage{221}\,--\,.
\end{bchapter}
\endbibitem

\bibitem[\protect\citeauthoryear{{Hurford}
  \textit{et~al.}}{2002}]{2002SoPh..210...61H}
\begin{barticle}
\bauthor{\bsnm{{Hurford}}, \binits{G.J.}},
\bauthor{\bsnm{{Schmahl}}, \binits{E.J.}},
\bauthor{\bsnm{{Schwartz}}, \binits{R.A.}},
\bauthor{\bsnm{{Conway}}, \binits{A.J.}},
\bauthor{\bsnm{{Aschwanden}}, \binits{M.J.}},
\bauthor{\bsnm{{Csillaghy}}, \binits{A.}},
\bauthor{\bsnm{{Dennis}}, \binits{B.R.}},
\bauthor{\bsnm{{Johns-Krull}}, \binits{C.}},
\bauthor{\bsnm{{Krucker}}, \binits{S.}},
\bauthor{\bsnm{{Lin}}, \binits{R.P.}},
\bauthor{\bsnm{{McTiernan}}, \binits{J.}},
\bauthor{\bsnm{{Metcalf}}, \binits{T.R.}},
\bauthor{\bsnm{{Sato}}, \binits{J.}},
\bauthor{\bsnm{{Smith}}, \binits{D.M.}}:
\byear{2002},
\batitle{{The RHESSI Imaging Concept}}.
\bjtitle{Solar Physics}
\bvolume{210},
\bfpage{61}\,--\,\blpage{86}.
\end{barticle}
\endbibitem

\bibitem[\protect\citeauthoryear{{Kosovichev}}{2011}]{2011ApJ...734L..15K}
\begin{barticle}
\bauthor{\bsnm{{Kosovichev}}, \binits{A.G.}}:
\byear{2011},
\batitle{{Helioseismic Response to the X2.2 Solar Flare of 2011 February 15}}.
\bjtitle{Astrophysical Journal}
\bvolume{734},
\bfpage{L15+}.
\end{barticle}
\endbibitem

\bibitem[\protect\citeauthoryear{Kosovichev and Zharkova}{1998}]{kz1998}
\begin{barticle}
\bauthor{\bsnm{Kosovichev}, \binits{A.G.}},
\bauthor{\bsnm{Zharkova}, \binits{V.V.}}:
\byear{1998},
\batitle{X-ray flare sparks quake inside the sun}.
\bjtitle{Nature}
\bvolume{393},
\bfpage{317}.
\end{barticle}
\endbibitem

\bibitem[\protect\citeauthoryear{Lindsey and Braun}{2000}]{lb2000}
\begin{barticle}
\bauthor{\bsnm{Lindsey}, \binits{C.}},
\bauthor{\bsnm{Braun}, \binits{D.C.}}:
\byear{2000},
\batitle{Basic principles of solar acoustic holography - (invited review)}.
\bjtitle{Solar Physics}
\bvolume{192},
\bfpage{261}\,--\,\blpage{284}.
\end{barticle}
\endbibitem

\bibitem[\protect\citeauthoryear{Mart\'inez-Oliveros, Moradi, and
  Donea}{2008}]{mo2008a}
\begin{barticle}
\bauthor{\bsnm{Mart\'inez-Oliveros}, \binits{J.C.}},
\bauthor{\bsnm{Moradi}, \binits{H.}},
\bauthor{\bsnm{Donea}, \binits{A.-C.}}:
\byear{2008},
\batitle{Seismic emissions from a highly impulsive m6.7 solar flare}.
\bjtitle{Solar Physics}
\bvolume{251},
\bfpage{613}\,--\,\blpage{626}.
\end{barticle}
\endbibitem

\bibitem[\protect\citeauthoryear{Mart\'inez-Oliveros
  \textit{et~al.}}{2007}]{mo2007}
\begin{barticle}
\bauthor{\bsnm{Mart\'inez-Oliveros}, \binits{J.C.}},
\bauthor{\bsnm{Moradi}, \binits{H.}},
\bauthor{\bsnm{Besliu-Ionescu}, \binits{D.}},
\bauthor{\bsnm{Donea}, \binits{A.-C.}},
\bauthor{\bsnm{Cally}, \binits{P.S.}},
\bauthor{\bsnm{Lindsey}, \binits{C.}}:
\byear{2007},
\batitle{From gigahertz to millihertz: A multiwavelength study of the
  acoustically active 14 august 2004 m7.4 solar flare}.
\bjtitle{Solar Physics}
\bvolume{245},
\bfpage{121}\,--\,\blpage{139}.
\end{barticle}
\endbibitem

\bibitem[\protect\citeauthoryear{Moradi \textit{et~al.}}{2007}]{metal2007}
\begin{barticle}
\bauthor{\bsnm{Moradi}, \binits{H.}},
\bauthor{\bsnm{Donea}, \binits{A.-C.}},
\bauthor{\bsnm{Lindsey}, \binits{C.}},
\bauthor{\bsnm{Besliu-Ionescu}, \binits{D.}},
\bauthor{\bsnm{Cally}, \binits{P.S.}}:
\byear{2007},
\batitle{Helioseismic analysis of the solar flare-induced sunquake of 2005
  january 15}.
\bjtitle{Monthly Notices of the Royal Astronomical Society}
\bvolume{374},
\bfpage{1155}\,--\,\blpage{1163}.
\end{barticle}
\endbibitem

\bibitem[\protect\citeauthoryear{{Schou}
  \textit{et~al.}}{2010}]{2010SoPh..tmp..177S}
\begin{botherref}
\oauthor{\bsnm{{Schou}}, \binits{J.}},
\oauthor{\bsnm{{Borrero}}, \binits{J.M.}},
\oauthor{\bsnm{{Norton}}, \binits{A.A.}},
\oauthor{\bsnm{{Tomczyk}}, \binits{S.}},
\oauthor{\bsnm{{Elmore}}, \binits{D.}},
\oauthor{\bsnm{{Card}}, \binits{G.L.}}:
2010,
{Polarization Calibration of the Helioseismic and Magnetic Imager (HMI) onborad
  the Solar Dynamics Observatory (SDO)}.
\textit{Solar Physics},
177\,--\,.
\end{botherref}
\endbibitem

\bibitem[\protect\citeauthoryear{{Sudol} and
  {Harvey}}{2005}]{2005ApJ...635..647S}
\begin{barticle}
\bauthor{\bsnm{{Sudol}}, \binits{J.J.}},
\bauthor{\bsnm{{Harvey}}, \binits{J.W.}}:
\byear{2005},
\batitle{{Longitudinal Magnetic Field Changes Accompanying Solar Flares}}.
\bjtitle{Astrophysical Journal}
\bvolume{635},
\bfpage{647}\,--\,\blpage{658}.
\end{barticle}
\endbibitem

\bibitem[\protect\citeauthoryear{Sudol and Harvey}{2005}]{sh2005}
\begin{barticle}
\bauthor{\bsnm{Sudol}, \binits{J.J.}},
\bauthor{\bsnm{Harvey}, \binits{J.W.}}:
\byear{2005},
\batitle{Longitudinal magnetic field changes accompanying solar flares}.
\bjtitle{Astrophysical Journal}
\bvolume{635},
\bfpage{647}\,--\,\blpage{658}.
\end{barticle}
\endbibitem

\bibitem[\protect\citeauthoryear{{Zharkov}
  \textit{et~al.}}{2011}]{2011ApJ...741L..35Z}
\begin{barticle}
\bauthor{\bsnm{{Zharkov}}, \binits{S.}},
\bauthor{\bsnm{{Green}}, \binits{L.M.}},
\bauthor{\bsnm{{Matthews}}, \binits{S.A.}},
\bauthor{\bsnm{{Zharkova}}, \binits{V.V.}}:
\byear{2011},
\batitle{{2011 February 15: Sunquakes Produced by Flux Rope Eruption}}.
\bjtitle{Astrophysical Journal}
\bvolume{741},
\bfpage{L35}.
\end{barticle}
\endbibitem

\bibitem[\protect\citeauthoryear{{Zharkova}}{2008}]{2008SoPh..251..665Z}
\begin{barticle}
\bauthor{\bsnm{{Zharkova}}, \binits{V.V.}}:
\byear{2008},
\batitle{{The Mechanisms of Particle Kinetics and Dynamics Leading to Seismic
  Emission and Sunquakes}}.
\bjtitle{Solar Physics}
\bvolume{251},
\bfpage{665}\,--\,\blpage{666}.
\end{barticle}
\endbibitem

\bibitem[\protect\citeauthoryear{{Zharkova}
  \textit{et~al.}}{2005}]{2005JGRA..11008104Z}
\begin{barticle}
\bauthor{\bsnm{{Zharkova}}, \binits{V.V.}},
\bauthor{\bsnm{{Zharkov}}, \binits{S.I.}},
\bauthor{\bsnm{{Ipson}}, \binits{S.S.}},
\bauthor{\bsnm{{Benkhalil}}, \binits{A.K.}}:
\byear{2005},
\batitle{{Toward magnetic field dissipation during the 23 July 2002 solar flare
  measured with Solar and Heliospheric Observatory/Michelson Doppler Imager
  (SOHO/MDI) and Reuven Ramaty High Energy Solar Spectroscopic Imager
  (RHESSI)}}.
\bjtitle{Journal of Geophysical Research (Space Physics)}
\bvolume{110},
\bfpage{8104}.
\end{barticle}
\endbibitem

\bibitem[\protect\citeauthoryear{Zharkova and Zharkov}{2007}]{zz2007}
\begin{barticle}
\bauthor{\bsnm{Zharkova}, \binits{V.V.}},
\bauthor{\bsnm{Zharkov}, \binits{S.I.}}:
\byear{2007},
\batitle{On the origin of three seismic sources in the proton-rich flare of
  2003 october 28}.
\bjtitle{Astrophysical Journal}
\bvolume{664},
\bfpage{573}\,--\,\blpage{585}.
\end{barticle}
\endbibitem

\end{thebibliography}

\end{article} 

\end{document}